\documentclass[prd,amsfonts,nofootinbib,showpacs]{revtex4}
\pdfoutput=1

\usepackage{amsmath}
\usepackage{amssymb}

\usepackage{ae} 
\usepackage{aecompl}
\usepackage{bm} 
\usepackage{graphicx}  
\usepackage{hyperref}

\usepackage{color}

\DeclareMathAlphabet{\mathsc}{OT1}{cmr}{m}{sc}
\newcommand {\ignore}[1]{}

\def\10{$SO(10)$}
\def\21{SU(2) $\otimes$ U(1) }

\def\422{$SU(4) \otimes SU(2) \otimes SU(2)$}
\def\321{SU(3) $\otimes$ SU(2) $\otimes$ U(1)}
\def\gsim{\raise0.3ex\hbox{$\;>$\kern-0.75em\raise-1.1ex\hbox{$\sim\;$}}}
\def\lsim{\raise0.3ex\hbox{$\;<$\kern-0.75em\raise-1.1ex\hbox{$\sim\;$}}}

\def\lsim{\raise0.3ex\hbox{$\;<$\kern-0.75em\raise-1.1ex\hbox{$\sim\;$}}}
\def\gsim{\raise0.3ex\hbox{$\;>$\kern-0.75em\raise-1.1ex\hbox{$\sim\;$}}}

\def \znbb {0\nu\beta\beta}

\newcommand{\AddrAHEP}{%
  AHEP Group, Institut de F\'{\i}sica Corpuscular --
  C.S.I.C./Universitat de Val{\`e}ncia \\
  Edificio Institutos de Paterna, Apt 22085, E--46071 Valencia, Spain}

\newcommand{\ba}{\begin{array}}
\newcommand{\ea}{\end{array}}
\relax
\def\321{$SU(3)\times SU(2)\times U(1)$}

\begin{document}

\begin{flushright}
\end{flushright}


\renewcommand{\Huge}{\Large} \renewcommand{\LARGE}{\Large}
\renewcommand{\Large}{\large} \def \znbb {$0\nu\beta\beta$ } \def \nbb
{$\beta\beta_{0\nu}$ } 
\title{Lepton flavor violation and non-unitary lepton mixing
in low-scale type-I seesaw}

\author{D.\,V.\,Forero}\email{dvanegas@ific.uv.es}
\author{S.~Morisi}\email{ morisi@ific.uv.es}
\author{M.\,T{\'o}rtola}\email{mariam@ific.uv.es} \author{J.~W.~F.~Valle}
\email{valle@ific.uv.es}

\affiliation{$^{1}$\AddrAHEP}

\date{\today}

\begin{abstract}

  Within low-scale seesaw mechanisms, such as the inverse and linear
  seesaw, one expects (i) potentially large lepton flavor violation
  (LFV) and (ii) sizeable non-standard neutrino interactions (NSI).
  We consider the interplay between the magnitude of non-unitarity
  effects in the lepton mixing matrix, and the constraints that follow
  from LFV searches in the laboratory.
  We find that NSI parameters can be sizeable, up to percent level in
  some cases, while LFV rates, such as that for $\mu\to e \gamma $,
  lie within current limits, including the recent one set by the MEG
  collaboration. As a result the upcoming long baseline neutrino
  experiments offer a window of opportunity for complementary LFV and
  weak universality tests.

\end{abstract}


\maketitle

\section{Introduction}
\label{intro}

One of the biggest challenges in particle physics is to unravel the
nature of the dimension five operator~\cite{weinberg:1980bf}
responsible for generating the pattern of neutrino masses and mixings
required to account for current oscillation
experiments~\cite{Schwetz:2011qt}. Within the seesaw
mechanism~\cite{th-talk2,valle:2006vb} this operator arises from the
tree-level exchange of heavy messenger particles.
For example, in the so-called type-I seesaw, these messengers are
three \321 singlet right-handed neutrinos, which must be super-heavy
in order to account for the observed smallness of neutrino masses.
However, since singlets carry no gauge-anomaly, their number is
theoretically unrestricted. Such generalized type-I seesaw
schemes~\cite{schechter:1980gr,schechter:1981cv} provide the
theoretical basis for the formulation of the inverse
seesaw~\cite{mohapatra:1986bd} and the linear seesaw
schemes~\cite{malinsky:2005bi}.
Both are capable of realizing the type-I seesaw mechanism at the TeV
scale and, as a result, these schemes open the possibility for novel
phenomena such as
\begin{enumerate}
\item lepton flavor and/or CP violating processes unsuppressed by
  neutrino
  masses~\cite{bernabeu:1987gr,gonzalez-garcia:1991be,branco:1989bn,rius:1989gk},
\item effectively non-unitary lepton mixing
  matrix~\cite{Antusch:2008tz,Malinsky:2009gw,Dev:2009aw} leading to
  non-standard effects in neutrino
  propagation~\cite{valle:1987gv,nunokawa:1996tg,Miranda:2004nb}.
\end{enumerate}
Both features arise from the non-trivial structure of the electroweak
currents in seesaw
schemes~\cite{schechter:1980gr,schechter:1981cv}~\footnote{They are
  generic in electroweak gauge models that mix fermions of different
  isospin in the weak currents~\cite{Lee:1977tib}.}.%
While their expected magnitude is negligible within the standard
type-I seesaw, it can be sizeable in low-scale seesaw mechanisms.

In this paper we analyse quantitatively the interplay between these
two classes of processes. More precisely we define reference
parameters describing the typical magnitude of non-standard neutrino
propagation effects and compare them with the constraints which arise
from the searches for lepton flavor violating (LFV)
processes~\cite{nakamura2010review,:2011ch,Ahmed:2001eh}. For
definiteness we focus on two simple realizations of low-scale seesaw,
namely the inverse and linear seesaw schemes. 
Non-unitarity effects may also arise from the charged lepton
sector~\cite{Ibanez:2009du}, leading also to lepton flavor violation.

We find that unitarity violation effects in neutrino propagation at
the percent level are consistent with the current bounds from lepton
flavor violation searches. Therefore their search at upcoming neutrino
oscillation facilities~\cite{bandyopadhyay:2007kx} opens a window of
opportunity to probe for new effects beyond the standard model.

The paper is organized as follows: in section \ref{sec:high-low-scale}
we review the type-I seesaw mechanisms, first the high-scale and then
the low-scale linear and inverse seesaw realizations, fixing the
notation and giving general expressions for the unitarity violating
parameters; in section~\ref{sec:unit-viol-magn} we study the
analytical expressions relating LFV decay branching ratios with the
non-standard pieces of the general seesaw lepton mixing matrix
characterizing inverse and linear seesaw schemes; in
section~\ref{sec:numerical-analysis} we describe the parameters used
in our numerical analysis, and in Sec.~\ref{sec:numerical-results} we
present our numerical results, both for normal and inverse neutrino
mass hierarchies.  Finally we give our conclusions.

\section{High and low-scale seesaw mechanisms}
\label{sec:high-low-scale}

The smallness of neutrino mass with respect to that of charged
fermions can be naturally accounted for within the seesaw mechanism,
in which the Standard Model is extended with extra \321
singlets~\cite{th-talk2,valle:2006vb}.
In this case the resulting neutrino mass matrix in general is a
$N\times N$ matrix with $N>3$. Here we analyse the structure of this
matrix in the high and low-scale type-I seesaw schemes.

\subsection{The standard type-I seesaw }
\label{sec:standard-type-i}

In general seesaw schemes the neutrino mass matrix $M_\nu$ can be
decomposed in sub-blocks involving the standard as well as singlet
neutrinos as follows
\begin{equation}\label{type-I}
M_\nu=\left(
\begin{array}{cc}
M_1&M_D\\
M_D^T &M_2
\end{array}
\right).
\end{equation}
in the basis $\nu,~\nu^c$, where the blocks $M_1\equiv M_L$ and
$M_2\equiv M_R$ are symmetric matrices.  Though the number of singlets
is arbitrary, here we take an equal number of SU(2) doublets and
singlets, and consider the simplest type-I seesaw, where no Higgs
triplet is present, so the upper left sub-matrix $M_1=0$ in
Eq.~(\ref{type-I})~\cite{schechter:1980gr}~\footnote{In this case
  light neutrinos get mass only as a result of the exchange of heavy
  gauge singlet fermions.}.
Neutrino masses arise by diagonalizing the matrix of
Eq.~(\ref{type-I}),
\begin{equation}\label{diagonalization}
 U^T\, M_\nu \, U= \text{real, diagonal}.
\end{equation}
through the transformation $U$ connecting the weak states to the light
and heavy mass eigenstates. We adopt a polar decomposition for $U$
where
\begin{equation}\label{expansion}
U =\exp(iH)\cdot V\,, \qquad
H= 
\left(
\begin{array}{cc}
0& S\\
S^\dagger & 0
\end{array}
\right),
\qquad V=
\left(
\begin{array}{cc}
V_1&0\\
0 & V_2
\end{array}
\right).
\end{equation}
so we have a power series expansion for Eq. (\ref{expansion}) given
as~\cite{schechter:1981cv}:
\begin{equation}\label{UI}
U =
 \left(
\begin{array}{cc}
\left(I-\frac{1}{2}S\,S^\dagger \right)\, V_1& i\,S\,V_2 \\
i\,S\,V_1 & \left(I-\frac{1}{2}S^\dagger \, S\right)\,V_2
\end{array}
\right) + O(\epsilon^3)
 \equiv
 \left(
 \begin{array}{cc}
 U_a & U_b \\
 U_c & U_d
\end{array}
\right),
\end{equation}
where we defined:
\begin{equation}
\epsilon\equiv M_D\,M_R^{-1}.
\end{equation}
Substituting Eqs.~(\ref{expansion}) and (\ref{type-I}) in
Eq.~(\ref{diagonalization}) one finds, from the requirement of
vanishing off-diagonal sub-blocks, that:
\begin{equation}\label{s-matrix}
 i\,S^*=-M_D\,M_R^{-1}.
\end{equation}
so that, using Eq. (\ref{s-matrix}) in Eq (\ref{expansion}) we
determine $U$ as:
\begin{equation}\label{u-expansion}
U=
\left(
\begin{array}{cc}
\left(I-\frac{1}{2}M_D^*(M_R^*)^{-1}\,M_R^{-1}M_D^T\right)V_1 & M_D^*(M_R^*)^{-1}\,V_2 \\
 -M_R^{-1}\,M_D^T \,V_1& \left(I-\frac{1}{2}M_R^{-1}\,M_D^T\,M_D^*\,(M_R^*)^{-1}\right)V_2
\end{array}
\right)+O(\epsilon^3),
\end{equation}
leading to an effective light neutrino mass matrix
\begin{equation}\label{normal}
m_\nu=-M_D M_R^{-1} M_D^T~.
\end{equation}
This is the so called type-I seesaw mechanism.
The smallness of neutrino masses follows naturally from the heaviness
of the $SU(2)_L$ singlet neutrino states $\nu_i$. Most seesaw
descriptions assume equal number of doublets and singlets, $n=m=3$.
However, since singlets carry no gauge-anomaly, their number is
arbitrary~\cite{schechter:1980gr,schechter:1981cv}.
In this paper we will consider not only the case $(n,m)=(3,3)$ just
described, but also the inverse and linear seesaw schemes, which
belong to the (3,6) class, see below.

\subsection{Inverse type-I seesaw}
\label{sec:inverse-seesaw}

As an alternative to the simplest \321 type-I seesaw model, it has
long been proposed extending the seesaw lepton content from $(3,3)$ to
$(3,6)$, by adding three extra $SU(2)$
singlets~\cite{mohapatra:1986bd}~\footnote{For simplicity we add the
  isosinglet pairs sequentially, though two pairs would suffice to
  account for the oscillations.}  $S_i$ charged under $U(1)_L$ global
lepton number the same way as the doublet neutrinos $\nu_i$, i.e.
$L=+1$.  After electroweak symmetry breaking one gets the mass matrix
\begin{equation} \label{inverse}
M_\nu=\left(
\begin{array}{ccc}
0&M_D&0\\
M_D^T & 0 &M\\
0&M^T& \mu
\end{array}
\right),
\end{equation}
in the basis $\nu$, $\nu^c,~S$, where the three $\nu^c_i$ have
$L=-1$. Note that $U(1)_L$ is broken only by the nonzero $\mu_{ij}
S_iS_j$ mass terms.

Generalizing the perturbative expansion method in
Ref.~\cite{schechter:1981cv} already used in the previous section one
finds that the mass matrix in Eq.~(\ref{inverse}) can be block
diagonalized as
\begin{equation}
\mathcal{U}^T\cdot M_\nu\cdot\mathcal{U}=\text{block diag}
\end{equation}
with
\begin{eqnarray}\label{u-bdiag}
&\mathcal{U}\approx
\left(
\begin{array}{ccc}
I & 0 & 0\\
0 & \frac{1}{\sqrt{2}}I & -\frac{1}{\sqrt{2}}I\\
0 & \frac{1}{\sqrt{2}}I & \frac{1}{\sqrt{2}}I
\end{array}
\right)
\left(
\begin{array}{ccc}
I-\frac{1}{2}S_1\,S_1^\dagger & 0 & iS_1\\
0 & I & 0\\
iS_1^\dagger & 0 & I-\frac{1}{2}S_1^\dagger \,S_1
\end{array}
\right)
\left(
\begin{array}{ccc}
I-\frac{1}{2}S_2\,S_2^\dagger & iS_2 & 0\\
iS_2\dagger & I-\frac{1}{2}S_2^\dagger \,S_2 & 0\\
0 & 0 & I
\end{array}
\right),&\nonumber\\
&
\approx
\left(
\begin{array}{ccc}
I & iS_2 & iS_1\\
-i\frac{1}{\sqrt{2}}\left(S_1^\dagger-S_2^\dagger \right) & \frac{1}{\sqrt{2}}I & -\frac{1}{\sqrt{2}}I\\
 i\frac{1}{\sqrt{2}}\left(S_1^\dagger+S_2^\dagger \right)& \frac{1}{\sqrt{2}}I & \frac{1}{\sqrt{2}}I
\end{array}
\right)+O(\epsilon^2)&
\end{eqnarray}
where $S_1$ and $S_2$ are $3\times 3$ matrices. In the limit $\mu \to 0$ we have  $S_1=S_2=S$ where
\begin{equation} \label{Uap}
iS^*=-\frac{1}{\sqrt{2}}m_D \,(M^T)^{-1}\sim \epsilon .
\end{equation}

With such matrix the light neutrino mass obtained after type-I seesaw
is
\begin{equation}\label{inv-mass}
m_\nu=M_DM^{T^{-1}}\mu M^{-1}M_D^T.
\end{equation}
Note that, in the limit as $\mu\to 0$ the lepton number symmetry is
recovered, making the three light neutrinos strictly massless. Thus
the smallness of neutrino mass follows in a natural way, in the sense
of 't Hooft \cite{thooft:1979}, as it is protected by $U(1)_L$.
One sees also that $\mathcal{U}$ consists of a maximal block rotation,
corresponding to the Dirac nature of the three heavy leptons made-up
of $\nu^c$ and $S$ in the limit as $\mu\to 0$, and two rotations
similar to Eq.\,(\ref{UI}) for the minimal type-I seesaw case
considered in the previous section.

Note also that the idea behind the so-called {\it inverse seesaw}
model can also be realized for other extended gauge groups
e.g. \cite{Wyler:1983dd,Akhmedov:1995vm,Barr:2005ss}.
Moreover, in specific models, the smallness of $\mu$ may be
dynamically generated \cite{Bazzocchi:2009kc}.

\subsection{Linear type-I seesaw}
\label{sec:linear-seesaw}

An alternative seesaw scheme that can also be realized at low-scale is
called the {\it linear seesaw}, and has been suggested as arising from
a particular $SO(10)$ unified model~\cite{malinsky:2005bi} (for other
possible constructions see~\cite{akhmedov:1995ip,Akhmedov:1995vm}).
Once the extended gauge structure breaks down to the standard \321 one
gets a mass matrix of the type
\begin{equation} \label{linear}
M_\nu=\left(
\begin{array}{ccc}
0&M_D&M_L\\
M_D^T & 0 &M\\
M_L^T&M^T& 0
\end{array}
\right),
\end{equation}
in the same basis $\nu, \nu^c, S$ used in
Sec.~\ref{sec:inverse-seesaw}.  Although theoretical consistency of
the model requires extra ingredients, such as Higgs scalars to
generate the $M_L$ and $M\equiv M_R$ entries, here we consider just
the simpler phenomenological scheme defined by the effective mass
matrix in Eq.~(\ref{linear}), as it suffices to describe the processes
we are interested in.

The block-diagonalization proceeds in a very similar way as to the
inverse seesaw case, in fact, for sufficiently small $M_L\,M^{-1}$ the
relations Eq. (\ref{u-bdiag}) and Eq.~(\ref{Uap}) are the same in both
schemes.  One finds that the effective light neutrino mass is now
given by
\begin{equation}\label{lin-mass}
m_\nu=M_D(M_LM^{{-1}})^T+ (M_LM^{-1})M_D^T.
\end{equation}
One sees that, in contrast to the ``usual'' seesaw relations for the
effective light neutrino mass, Eqs.~(\ref{normal}) and
(\ref{inv-mass}), the formula in Eq.~(\ref{lin-mass}) is linear in the
Dirac neutrino Yukawa couplings, hence the name {\sl linear seesaw}.

Notice also that the lepton number, defined as in the previous model,
is broken only by the terms $M_L\,\nu^c \, S$.  As a result one sees
that, in the limit as $M_L\to 0$ the lepton number symmetry is
recovered, making the three light neutrinos strictly massless. Again,
as in the previous case, the smallness of neutrino mass follows in a
natural way~\cite{thooft:1979}, as it is protected by $U(1)_L$.

\section{ Unitarity violation and the magnitude of lepton flavor
  violation}
\label{sec:unit-viol-magn}

The effective lepton mixing matrix $K_{i\alpha}$ characterizing the
charged current weak interaction of mass-eigenstate neutrinos in any
type of seesaw model has been fully characterized in
Ref.~\cite{schechter:1980gr}. It can be expressed in rectangular form
\begin{equation}
\mathcal{L}\supset i\frac{g}{\sqrt{2}}W_\nu \overline{l}_bK_{b\alpha}\gamma_\mu \nu_{\alpha L}+h.c.~,
\end{equation}
where
\begin{equation}\label{kappa}
K_{b\alpha}=\sum_{c=1}^n \Omega^*_{cb}U_{c\alpha}~,
\end{equation}
where $\Omega$ is the 3 by 3 unitary matrix that diagonalizes the
charged lepton mass matrix, while $U$ is the unitary matrix that
diagonalizes the (higher-dimensional) neutrino mass matrix
characterizing the type-I seesaw mechanism of interest.  We may write
the $K$ matrix as follows
\begin{equation}
K=\left(
K_{L},K_{H}
\right)~,
\end{equation}
where $K_{L}$ is a 3 by 3 matrix and $K_H$ is a 3 by 6 matrix. While
the rows of the $K$ matrix are unit vectors, since $K\cdot K^\dagger =
I$, the blocks $K_L$ and $K_H$ are not unitary. 

For our purposes we can take the charged lepton mass matrix in
diagonal form~\footnote{This may be automatic in the presence of
  suitable discrete flavor symmetries as in \cite{Hirsch:2009mx}.} so
that $\Omega \to 1$. From Eq. (\ref{u-expansion}) we obtain:
\begin{equation}\label{unit-dev}
\begin{split}
 K_L&=\left(1-\frac{1}{2}M_D^*(M^*)^{-1}(M)^{-1}M_D^T\right)V_1~,\\
 K_H&=\left(M_D^*(M^*)^{-1}\right)V_2.
\end{split}
\end{equation}

In order to establish a simple comparison with recent literature we
parametrize the deviation from unitarity
as~\cite{Hettmansperger:2011bt}
\begin{equation}\label{unit-dev-eta}
K_L \equiv (1-\eta)V_1.
\end{equation}
Then for the simplest high-scale type-I see-saw, the deviation from
unitarity characterizing the mixing of light neutrinos,
Eq.\,(\ref{unit-dev}), is given by
\begin{equation}\label{eta}
 \eta \sim \frac{1}{2}\epsilon^* \epsilon^T\approx \frac{1}{2}M_D^*(M^*)^{-1}(M)^{-1}M_D^T.
\end{equation}
Barring {\sl ad hoc} fine-tuning, it follows that for this case one
expects negligible deviation from unitarity, namely $\epsilon\approx
10^{-10}$ and so $\eta\approx 10^{-20}$.

From now on we focus on the low-scale type-I seesaw schemes discussed
in Secs~\ref{sec:inverse-seesaw} and ~\ref{sec:linear-seesaw}, inverse
and linear seesaw, respectively.
Generalizing the above discussion to these cases one finds 
\begin{equation}
 K_L=\left[I-\frac{1}{2}\left( M_D^*((M^T)^*)^{-1}(M)^{-1}M_D^T\right) \right]V_1
\end{equation}
Hence the parameters characterizing the deviation from unitarity
analogous to Eq. (\ref{eta}) are given by~\footnote{For a recent study
  of unitarity violation in seesaw schemes see, for instance,
  Ref.~\cite{Hettmansperger:2011bt}.},
\begin{equation}\label{eta-inverse}
 \eta^{I,L} \approx \frac{1}{2}\left(M_D^*((M^T)^*)^{-1}(M)^{-1}M_D^T\right),
\end{equation}
which holds for both the type-I inverse and linear seesaw
mechanisms. These parameters characterize the corresponding unitarity
deviation in the light-active $3\times 3$ sub-block of the lepton
mixing matrix.

\begin{figure}
\begin{center}
\includegraphics[angle=0,height=4cm,width=0.24\textwidth]{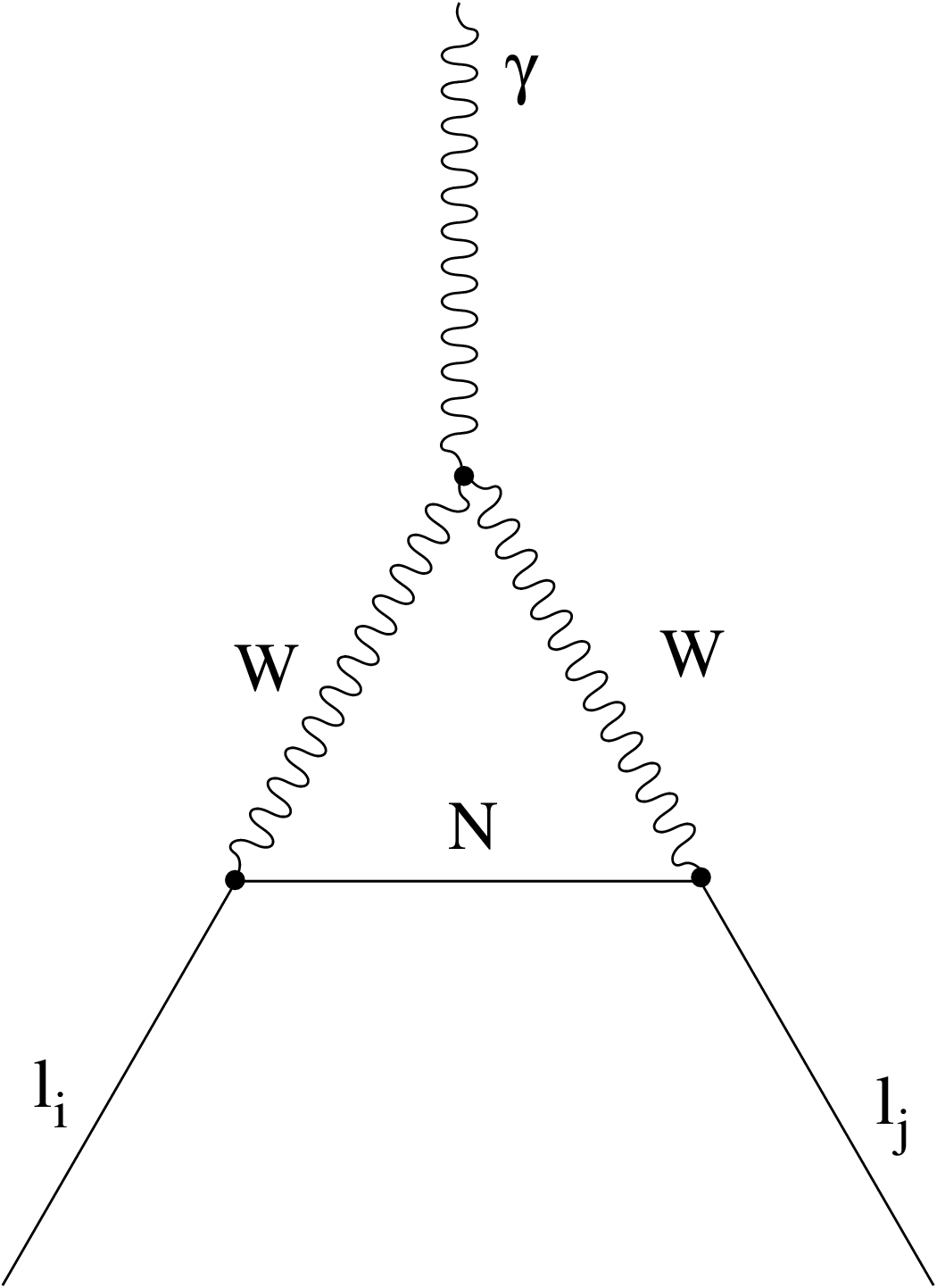}
\caption{Feynman graphs for $\mu \to e \gamma$ decay in 
  seesaw models.}
\label{ref:feynmangraphs}
\end{center}
\end{figure}

\begin{figure}[!h]
\begin{center}
\includegraphics[angle=0,height=7cm,width=0.48\textwidth]{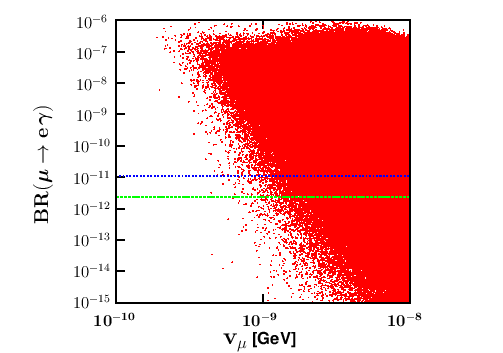}
\includegraphics[angle=0,height=7cm,width=0.48\textwidth]{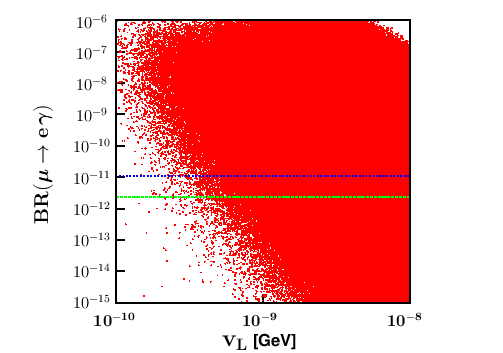}
\caption{$Br(\mu\to e \gamma)$ versus the lepton number violation
  scale: $v_\mu$ for the inverse seesaw (left panel), and $v_L$ for
  the linear seesaw (right panel). In both cases one assumes normal
  hierarchy and the parameters are varied as explained in
  Sec.~\ref{sec:numerical-analysis}. We also indicate the limits from
  the MEGA collaboration~\cite{Ahmed:2001eh} (upper line) and the
  recent limit from MEG~\cite{:2011ch} (lower line).}
\label{fig:MEGBR}
\end{center}
\end{figure}

We now turn to the lepton flavor violating processes that would be
induced at one loop in type-I seesaw models, as a result of the mixing
of SU(2) doublet neutrinos with singlet neutral heavy leptons. The
latter breaks the Glashow-Illiopoulos-Maiani cancellation
mechanism~\cite{glashow:1970gm}, enhancing the rates for the
loop-induced lepton flavor violating processes illustrated in
Fig.~\ref{ref:feynmangraphs}.
The $l_i \to l_j \gamma $ decay process is induced through the
exchange of the nine neutral leptons coupled to the charged leptons in
the charged current, namely the three light neutrinos as well as the
six sub-dominantly coupled heavy
states~\cite{bernabeu:1987gr,Ilakovac:1994kj,deppisch:2004fa,He:2002pva,Lavoura:2003xp}.
The resulting decay branching ratio is given by
\begin{equation}\label{Br}
Br(l_i\to l_j \gamma)= \frac{\alpha^3_Ws_W^2}{256 \pi^2}\frac{m_{l_i}^5}{M_W^4}
\frac{1}{\Gamma_{l_{i}}}|G_{ij}^W|^2~,
\end{equation}
where we use the explicit analytic form of the loop-functions~\cite{He:2002pva},
\begin{equation}\label{def:G}
\begin{array}{l}
G_{ij}^W=\sum_{k=1}^9 K^*_{ik} K_{jk} G_\gamma^W\left(\frac{m^2_{N_k}}{M_W^2}\right)~,\\
G_\gamma^W(x)=\frac{1}{12(1-x)^4}(10-43x+78x^2-49x^3+18x^3\ln{x}+4x^4)~\\
\end{array}
\end{equation}
and is presented in Fig.~\ref{fig:MEGBR} for each case. The difference
between the two models follows from the different dependence with the
lepton number violating parameters that characterize these two
low-energy type-I seesaw realizations.

One sees that the branching rations may easily exceed current limits.
This reflects an important feature of low-scale seesaw models, namely,
that lepton flavor violation as well as leptonic CP violation proceed
even in the limit of massless
neutrinos~\cite{bernabeu:1987gr,gonzalez-garcia:1991be,branco:1989bn,rius:1989gk}.
Unsuppressed by the smallness of neutrino mass, the expected rates are
sizeable. The corresponding radiative LFV decays $\tau\to e \gamma$
and $\tau\to \mu \gamma$ are not as constraining as $\mu\to e
\gamma$. Similarly, the fact that the neutral current couplings of
charged leptons is flavor diagonal implies that the decay processes
$l_i \to l_jl_kl_r$ are suppressed by a factor of $\alpha_{QED}$ with
respect to the radiative processes, hence less restrictive.

Another very important lepton flavor violating process is mu-e
conversion in nuclei, which arises both from short range
(non-photonic) as well as long range (photonic)
contributions. Explicit calculations~\cite{deppisch:2005zm} indicate
that, given the nuclear form factors, one finds that current mu-e
conversion sensitivities are effectively lower than those of the $\mu
\to e \gamma$ decay. However the upcoming generation of nuclear
conversion experiments aims at substantial improvement.

\section{Numerical analysis}
\label{sec:numerical-analysis}

In order to perform our numerical calculations it is convenient to
generalize the Casas-Ibarra parametrization~\cite{casas:2001sr} to the
inverse and linear type-I seesaw schemes. For simplicity we will
assume real lepton Yukawa couplings and mass entries.

\subsection{Inverse type-I Seesaw}
\label{sec:inverse-type-i}

First note that one has always the freedom to go to the basis where
the $3\times 3$ gauge-singlet block $M$ is taken diagonal.
For real $m_D$ matrix elements, we have in total $18$ parameters, nine
characterizing $m_D$, three characterizing $M$, plus six from the
$\mu$ matrix.

The Dirac neutrino mass matrix may be rewritten as
\begin{equation}\label{mdirac-inv}
m_D=V_1\, \text{diag}(\sqrt{\tilde{m}_i}) \, R^T \, {(\sqrt{\mu})}^{-1}~\text{diag} (M^T_i)
\end{equation}
where $V_1$ is (approximately) the mixing matrix determined in
oscillation experiments~\cite{Schwetz:2011qt}, $\tilde{m}_i$ are the
three light neutrino masses. On the other hand the arbitrary real
orthogonal $3\times 3$ matrix $R$ and the arbitrary $3\times 3$ real
matrix $M$ are parameters characterizing the model.
This parametrization for the inverse type-I seesaw is similar to that
given in Ref.~\cite{deppisch:2004fa}.

In order to further reduce the number of degrees of freedom, we will
make the ``minimal flavor violation hypothesis''~\footnote{This
  simplifying assumption suffices to illustrate the points made in
  this paper. For alternative minimal flavor violation definitions see
  Refs.~\cite{D'Ambrosio:2002ex,Cirigliano:2005ck,Alonso:2011jd}.}
which consists in assuming that flavor is violated only in the
``standard'' Dirac Yukawa coupling. Under this simplification the
$3\times 3$ matrix $\mu$ must be also diagonal, reducing the parameter
count from a total of $18$ down to $15$. These include the three light
neutrino masses, and the three neutrino mixing angles contained in
$V_1$.  Next come the nine model-defining parameters, that may be
taken as three parameters from the $R$ matrix, three from the $\mu$
matrix, plus three parameters characterizing $M$.

We have performed a scan at $3\sigma$ over the lightest mass and
oscillation parameters, the three angles and the three masses in $V_1$
and $\tilde{m}_i$, respectively. For the scan over oscillation
parameters we have used the $3\sigma$ determinations given
in~\cite{Schwetz:2011qt}, and for the lightest mass parameter we took
the cosmological bound from \cite{Fogli:2008}.
We parametrize the real orthogonal matrix $R$ as a product of three
rotations, marginalizing over the three angles from $0-2\pi$ values.
 
We have also fixed the upper value of the Dirac mass matrix to
$(m_D)_{ij}<175\,GeV$ to be consistent with perturbativity of the
theory.  The remaining six free parameters, from $diag\{\mu_{ii}\}$
and $diag\{M_{ii}\}$ matrices, are scanned as a perturbation from the
identity matrix in the following way:
\begin{equation}\label{perturbation}
\begin{split}
\mu_{ii}&=v_\mu \, \left(1+\varepsilon_{ii}\right) \\
 M_{ii}&=v_M \, \left(1+\varepsilon^\prime_{ii}\right),
\end{split}
\end{equation}
where $|\varepsilon|\sim 5\times10^{-1}$. The parameter $v_M$ setting
the $M$-scale was fixed to $1\,TeV$, while $v_\mu$ scale was scanned
in the range $(0.1-10)\,eV$. The two scales $v_{\mu,M}$ are consistent
with the observed neutrino masses.

\subsection{Linear type-I Seesaw}
\label{sec:linear-type-i}

Similarly, for the linear seesaw case we can parametrize Dirac
neutrino mass matrix as follows,
\begin{equation}\label{mdirac-lin}
m_D=V_1 \text{diag}\{\sqrt{m_i}\} \, A^T \text{diag}\{\sqrt{m_i}\} \, V_1^T \left(M_L^T\right)^{-1} \,M^T,
\end{equation}
where $A$ has the following general form:
\begin{equation}\label{A}
\left(
\begin{array}{ccc}
\frac{1}{2} & a & b\\
-a & \frac{1}{2}& c\\
-b & -c  & \frac{1}{2}
\end{array}
\right),
\end{equation}
with $a,b,c$ real numbers.
In this case $M_L$ and $M\equiv M_R$ are general real matrices. One
can always go to a basis where one of them is diagonal, for example
$M_L$, reducing the total number of model parameters to 21.

In order to further reduce the number of degrees of freedom, we make a
similar ``minimal flavor violation hypothesis'' to this scheme too,
namely, we choose the $M$ matrix also to be diagonal, reducing from 21
to 15 parameters.
The most general Dirac mass matrix is parametrized as in
Eq.~(\ref{mdirac-lin}). Hence we have a model with $15$ parameters
that may be taken as three parameters from $A$ matrix, three from the
$M_L$ matrix, three from the matrix $M$, in addition to the three
light neutrino masses and three mixing parameters.

As for the inverse type-I seesaw case, we could do the scan over the
light neutrino mass and oscillation parameters and the 9 free
parameters.  The difference in this case respect to the inverse, is
the structure of the $A$ matrix in Eq.~(\ref{A}).  We scan over the
$A$ matrix parameters in the form:
\begin{equation}
A_{ij} \, \, \epsilon \, (0-10^{-2}),
\end{equation}
and now we define $M_{Lii}$ in analogy to $M_{ii}$ in
Eq.~(\ref{perturbation}) and we vary $v_L=(0.01-10)\,eV$.

\section{Numerical results}
\label{sec:numerical-results}

Low-scale seesaw schemes lead to sizeable rates for lepton flavor
violating processes as well as to non-standard effects in neutrino
propagation associated to non-unitary lepton mixing matrix.
In this section we will quantify the interplay between these, more
precisely, between the branching ratio of Eq. (\ref{Br}) in the
low-scale type-I seesaw schemes considered here and the magnitude of
the unitarity deviation defined in Eq.~(\ref{eta}), taking into
account Eqs.~(\ref{mdirac-inv}) and (\ref{mdirac-lin}) and the
requirement of acceptable light neutrino masses. For definiteness we
assume leptonic CP conservation so that all lepton Yukawa couplings
and mass entries are real.

\begin{figure*}[!h]
\centerline{
\includegraphics[scale=0.35]{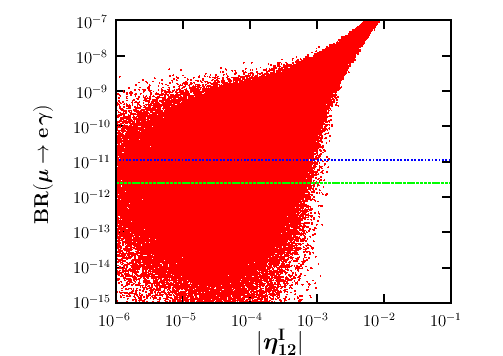}
\includegraphics[scale=0.35]{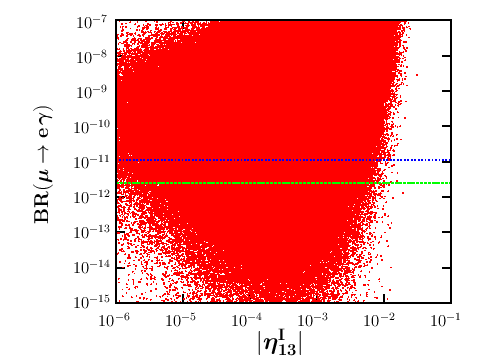}
\includegraphics[scale=0.35]{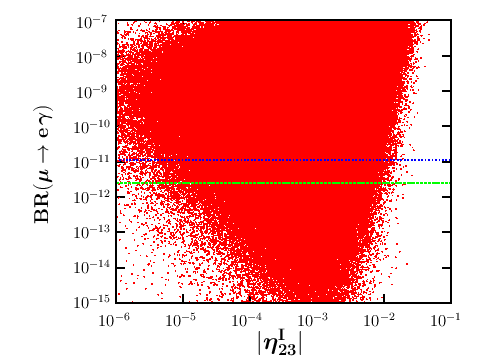}
}
\caption{\label{fig-inverse-br-NH-me} Branching for the process $\mu
  \to e \gamma$ in type-I inverse seesaw scheme with normal hierarchy
  (NH).  We scan over the light neutrino mass and mixing parameters at
  3 $\sigma$, and over the model parameters, fixing $v_M$ at 1~$TeV$
  and varying the $v_\mu$ scale from $1\times 10^{-10}\,GeV$ to $1
  \times 10^{-8}\,GeV$.  We also indicate the limits from the MEGA
  collaboration (upper line) and the recent limit from MEG (lower
  line).  }
\end{figure*}
\begin{figure*}[!h]
\centerline{
\includegraphics[scale=0.35]{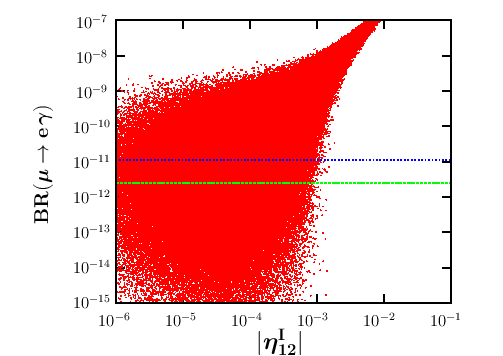}
\includegraphics[scale=0.35]{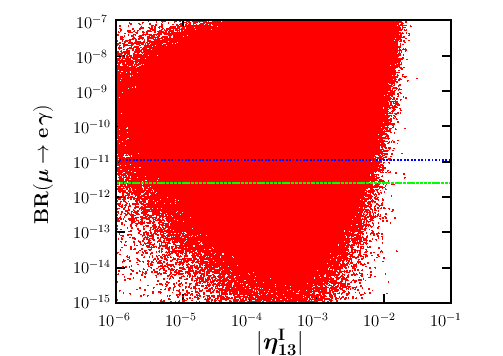}
\includegraphics[scale=0.35]{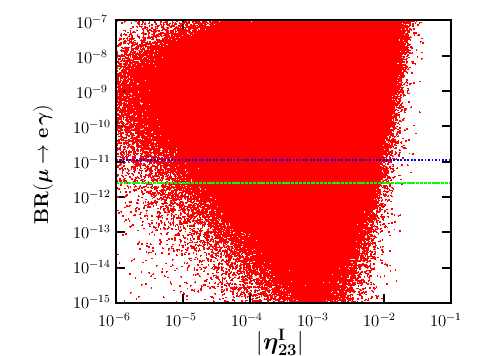}
}
\caption{\label{fig-inverse-br-IH-me} 
  Same as Fig.~\ref{fig-inverse-br-NH-me} for the Inverted Hierarchy
  (IH) case.  }
\end{figure*}

We have computed the branching ratio (BR) for the charged lepton
flavor violating radiative processes using Eq.~(\ref{Br}), accurate to
order $O(\epsilon^3)$ in the neutrino diagonalizing matrix, and
displayed the degree of correlation of these observables with the
corresponding unitarity violating parameters $|\eta_{ij}|$.

In Fig. \ref{fig-inverse-br-NH-me} and Fig. \ref{fig-inverse-br-IH-me}
we show the results for the inverse type-I seesaw scheme with Normal
(NH) and Inverted (IH) neutrino mass hierarchy for the process $\mu
\to e \gamma$, respectively.  The points result from a scan over the
light neutrino mass and mixing parameters at 3 $\sigma$, together with
the scan over the 9 free model parameters defined above.

\begin{figure*}[!htb]
\centerline{
\includegraphics[scale=0.35]{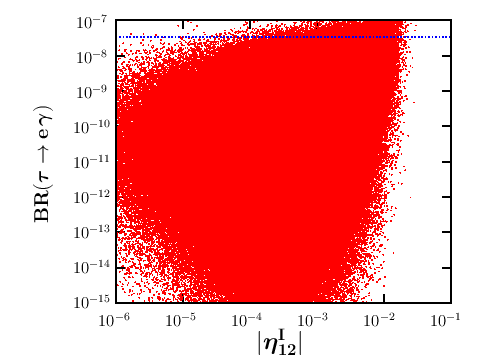}
\includegraphics[scale=0.35]{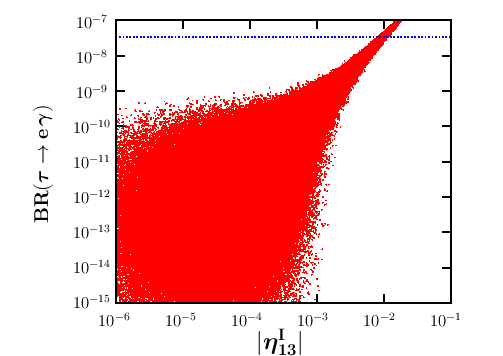}
\includegraphics[scale=0.35]{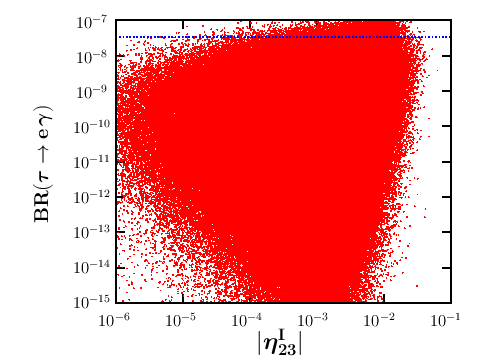}
}
\caption{\label{fig-inverse-br-NH-te} $\tau \to e \gamma$ branching
  ratio in the inverse type-I seesaw with Normal Hierarchy (NH).  The
  scan is performed as in Fig.~\ref{fig-inverse-br-NH-me} and the
  indicated limit is from the PDG~\cite{nakamura2010review}.  }
\end{figure*}

\begin{figure*}[!htb]
\centerline{
\includegraphics[scale=0.35]{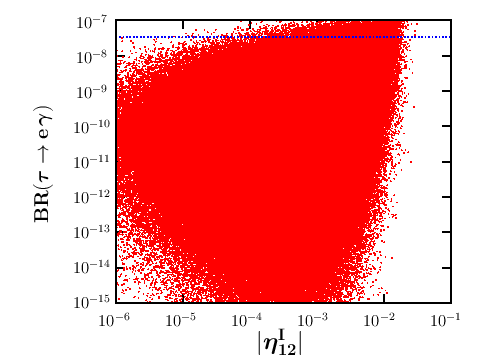}
\includegraphics[scale=0.35]{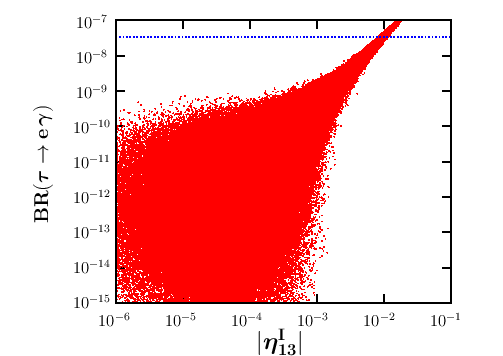}
\includegraphics[scale=0.35]{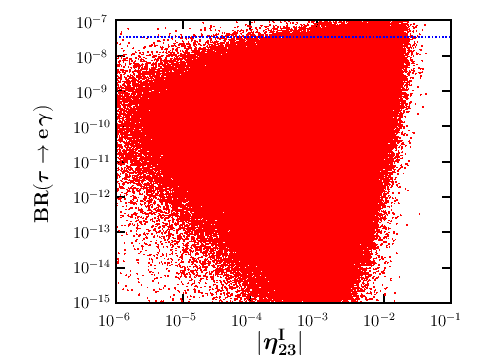}
}
\caption{\label{fig-inverse-br-IH-te} Same as Fig.~\ref{fig-inverse-br-NH-te}
for  Inverted Hierarchy (IH).
}
\end{figure*}

In Fig.~\ref{fig-inverse-br-NH-te} and Fig.~\ref{fig-inverse-br-IH-te}
we show the corresponding results for the $\tau \to e \gamma$
branching ratio in the inverse type-I seesaw scheme with Normal
Hierarchy (NH) and Inverted Hierarchy (IH), respectively.

\begin{figure*}[!htb]
\centerline{
\includegraphics[scale=0.35]{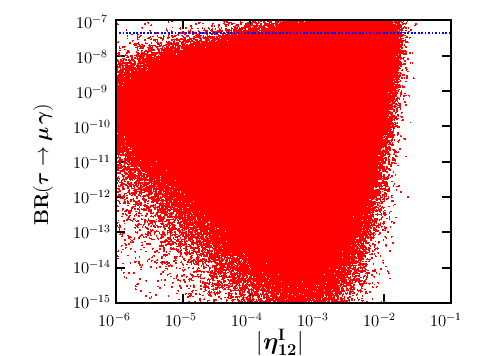}
\includegraphics[scale=0.35]{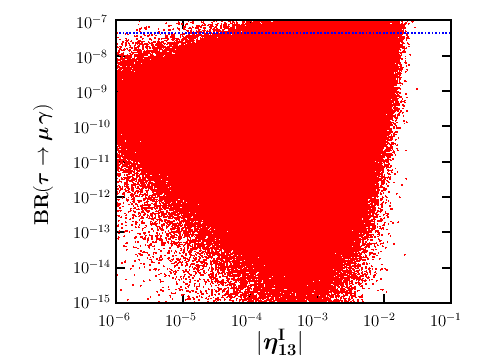}
\includegraphics[scale=0.35]{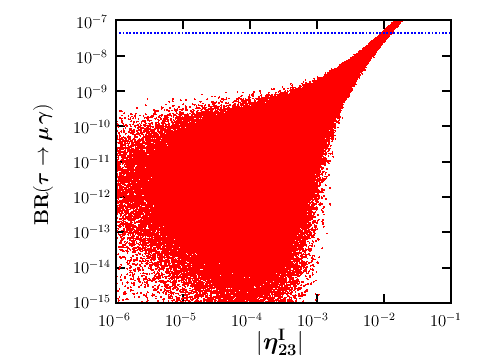}
}
\caption{\label{fig-inverse-br-NH-tm} 
  $\tau \to \mu \gamma$ branching ratio in the inverse type-I seesaw
  with Normal Hierarchy (NH). The scan is performed as in
  Fig.~\ref{fig-inverse-br-NH-me} and the indicated limit is from the
  PDG~\cite{nakamura2010review}.  }
\end{figure*}

\begin{figure*}[!htb]
\centerline{
\includegraphics[scale=0.35]{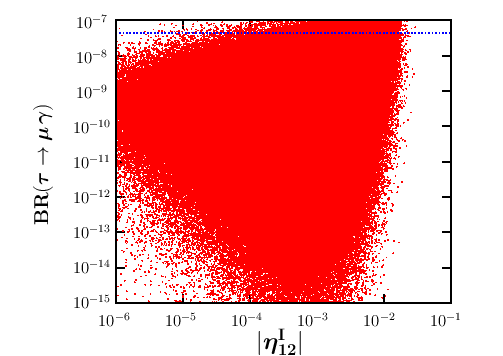}
\includegraphics[scale=0.35]{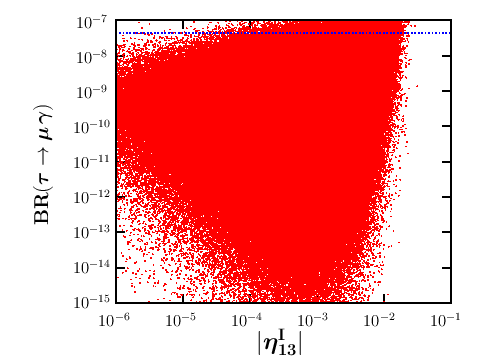}
\includegraphics[scale=0.35]{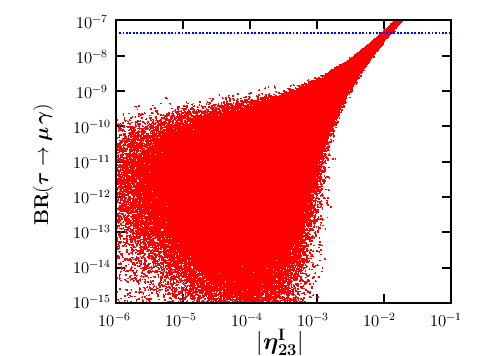}
}
\caption{\label{fig-inverse-br-IH-tm} 
Same as Fig.~\ref{fig-inverse-br-NH-tm} for  Inverted Hierarchy (IH).
}
\end{figure*}

In Fig. \ref{fig-inverse-br-NH-tm} and Fig. \ref{fig-inverse-br-IH-tm}
we show the corresponding results for the process $\tau \to \mu
\gamma$ within the inverse type-I seesaw, for  Normal Hierarchy (NH) 
and Inverted Hierarchy (IH) neutrino spectra, respectively.

\begin{figure*}[!htb]
\centerline{
\includegraphics[scale=0.35]{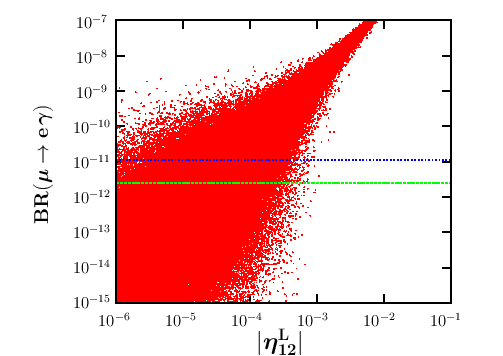}
\includegraphics[scale=0.35]{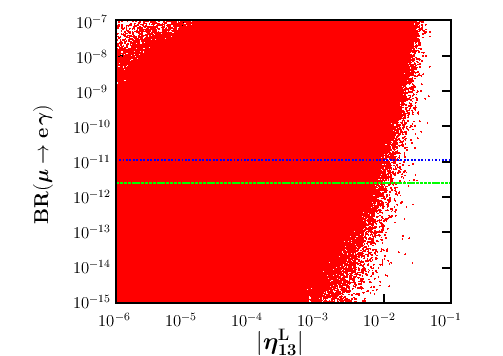}
\includegraphics[scale=0.35]{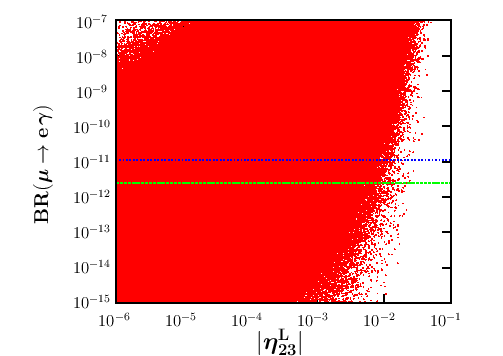}
}
\caption{\label{fig-linear-br-NH-me} Branching for the process $\mu
  \to e \gamma$ in type-I linear seesaw with Normal Hierarchy (NH).
  We scan over the parameters as in Fig.~\ref{fig-inverse-br-NH-me}.
  The limits from the MEGA and MEG collaborations are indicated by the
  upper and lower horizontal lines.  }
\end{figure*}

\begin{figure*}[!htb]
\centerline{
\includegraphics[scale=0.35]{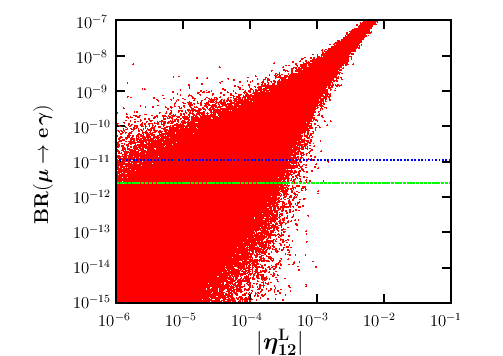}
\includegraphics[scale=0.35]{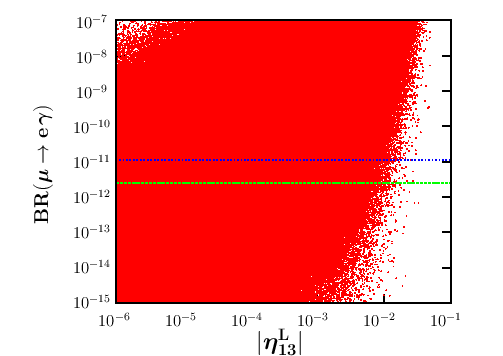}
\includegraphics[scale=0.35]{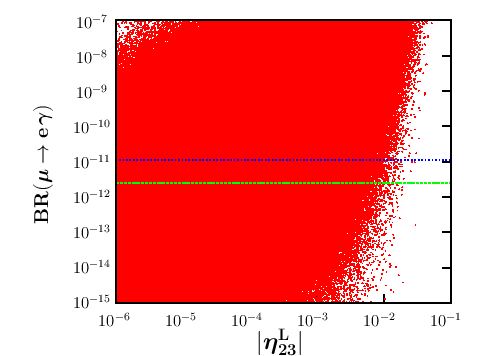}
}
\caption{\label{fig-linear-br-IH-me} 
Same as in Fig.~\ref{fig-linear-br-NH-me} for Inverted Hierarchy (IH).
}
\end{figure*}

Now we turn to the linear type-I seesaw scheme.  In
Fig.~\ref{fig-linear-br-NH-me} and Fig.~\ref{fig-linear-br-IH-me} we
show our results for the branching ratio for the process $\mu \to e
\gamma$ in such linear seesaw for Normal Hierarchy (NH) and Inverted
Hierarchy (IH), respectively. The points are obtained through a scan
over the neutrino oscillation parameters, as well as the free model
parameters, as already described.

\begin{figure*}[!htb]
\centerline{
\includegraphics[scale=0.35]{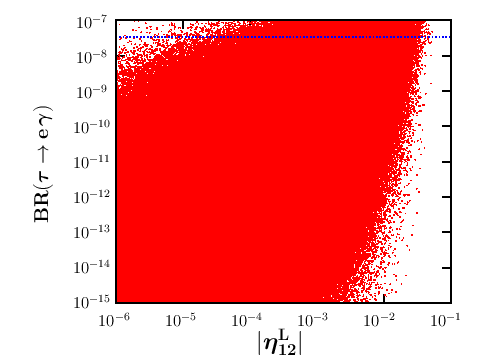}
\includegraphics[scale=0.35]{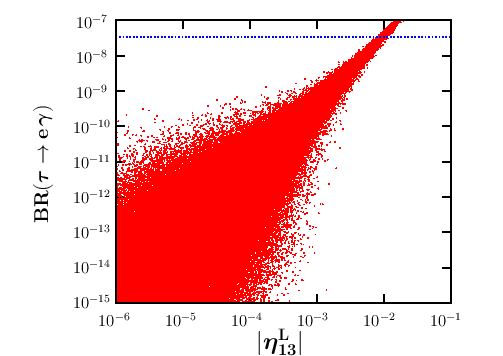}
\includegraphics[scale=0.35]{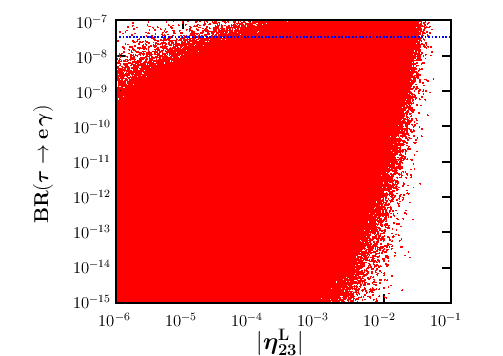}
}
\caption{\label{fig-linear-br-NH-te} $\tau \to e \gamma$ branching
  ratio, in Linear type-I seesaw for Normal Hierarchy (NH). Parameters
  scanned as in Fig.~\ref{fig-inverse-br-NH-me} and the indicated
  limit is from the PDG~\cite{nakamura2010review}.  }
\end{figure*}

\begin{figure*}[!htb]
\centerline{
\includegraphics[scale=0.35]{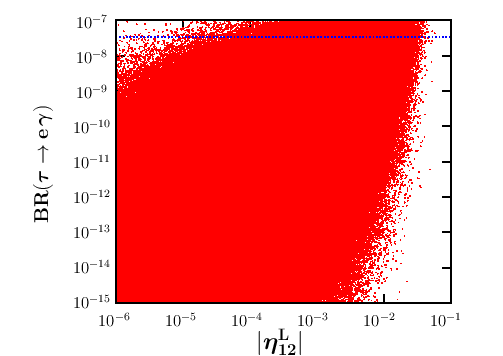}
\includegraphics[scale=0.35]{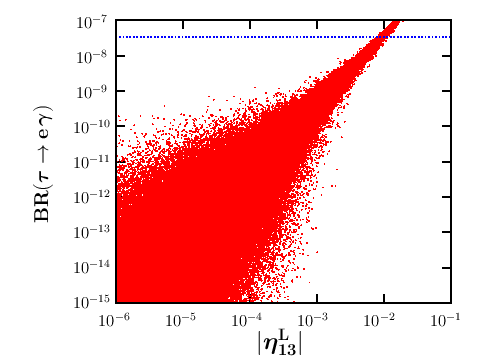}
\includegraphics[scale=0.35]{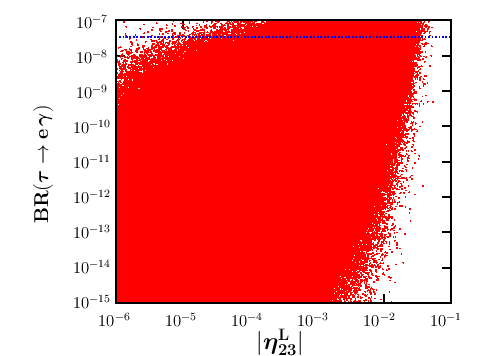}
}
\caption{\label{fig-linear-br-IH-te} 
Same as Fig.~\ref{fig-linear-br-NH-te} for Inverted Hierarchy (IH).
}
\end{figure*}

\begin{figure*}[!htb]
\centerline{
\includegraphics[scale=0.35]{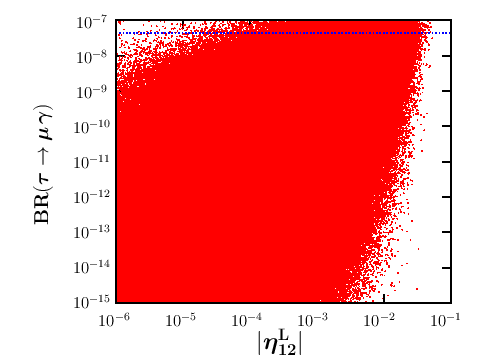}
\includegraphics[scale=0.35]{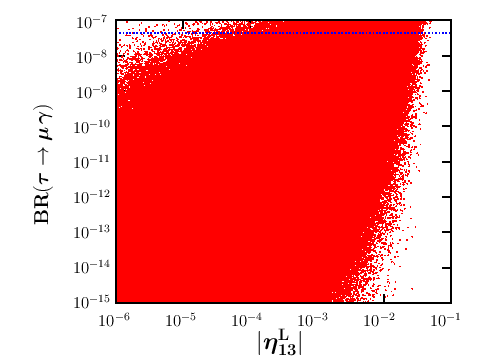}
\includegraphics[scale=0.35]{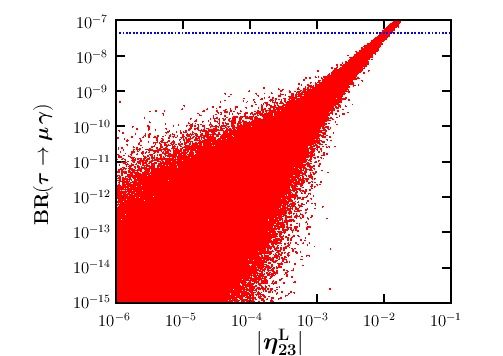}
}
\caption{\label{fig-linear-br-NH-tm} $\tau \to \mu \gamma$ branching
  ratio, in Linear type-I seesaw for Normal Hierarchy (NH). Parameters
  scanned as in Fig.~\ref{fig-inverse-br-NH-me} and the indicated
  limit is from the PDG~\cite{nakamura2010review}.  }
\end{figure*}

\begin{figure*}[!htb]
\centerline{
\includegraphics[scale=0.35]{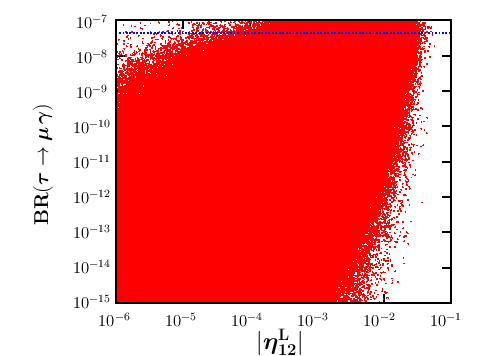}
\includegraphics[scale=0.35]{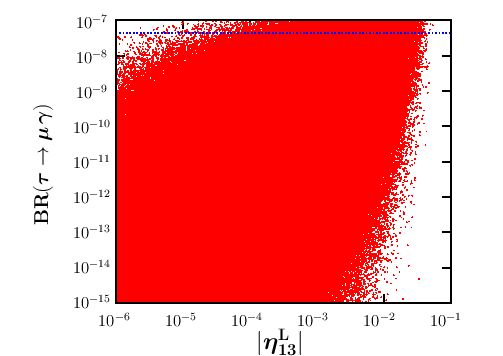}
\includegraphics[scale=0.35]{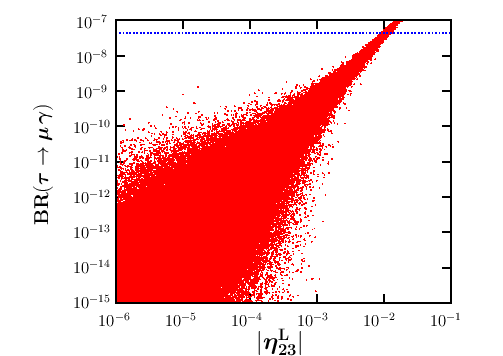}
}
\caption{\label{fig-linear-br-IH-tm} 
Same as Fig.~\ref{fig-linear-br-NH-te} for Inverted Hierarchy (IH).
}
\end{figure*}

In Fig.~\ref{fig-linear-br-NH-te} and Fig.~\ref{fig-linear-br-IH-te}
we show our results for the $\tau \to e \gamma$  branching ratio, 
for the Normal Hierarchy (NH) and Inverted Hierarchy (IH)
case, respectively.
Finally, in Fig.~\ref{fig-linear-br-NH-tm} and Fig.~\ref{fig-linear-br-IH-tm}
we present the corresponding results for the   $\tau \to \mu \gamma$  process.

These results are summarized in table~\ref{tab-results2}. One sees
that the magnitude of non-unitarity effects in the lepton mixing
matrix can reach up to percent level is not in conflict with
the constraints that follow from lepton flavor violation searches in
the laboratory.  Given the large - TeV scale - assumed masses of the
singlet ``right-handed'' neutrinos there are no direct search
constraints~\cite{dittmar:1989yg,Kovalenko:2009td,Atre:2009rg}.  The
main factor limiting the magnitude of non-unitarity effects then
becomes the weak universality constraints. As expected, there is 
stronger degree of correlation between $\mu  \to e \gamma$ and 
$\eta_{12}$ than other $\eta$'s, or between $\tau  \to e \gamma$ 
and $\eta_{13}$ than others, etc. As a result in these cases one 
obtains the strongest restriction on unitarity violation.

\begin{table}[!htb]
  \centering
\begin{tabular}{c| c c |c c| c c} \hline \hline
 Process & $\mu \to e \gamma$ & & $\tau \to e \gamma$ & & $\tau \to \mu \gamma$\\ \hline
 Hierarchy & NH & IH & NH & IH & NH & IH	\\ \hline
$|\eta_{12}^I|<$  & $ 1.4\times 10^{-3}$ & $ 1.4\times 10^{-3}$ & $2.8\times 10^{-2}$ & $2.8\times 10^{-2}$ & $2.8\times 10^{-2}$ & $2.8\times 10^{-2}$\\
 $|\eta_{13}^I|<$  & $ 2.0\times 10^{-2}$ & $2.1(1.6)\times 10^{-2}$ & $1.1\times 10^{-2}$ & $1.1\times 10^{-2}$ & $3.1\times 10^{-2}$ & $3.2\times 10^{-2}$\\
 $|\eta_{23}^I|<$  & $2.7(2.1)\times 10^{-2}$ & $2.5(1.9)\times 10^{-2}$ & $6.4\times 10^{-2}$ & $4.3\times 10^{-2}$ & $1.2\times 10^{-2}$ & $1.2\times 10^{-2}$\\ \hline
 $|\eta_{12}^L|<$  & $11.0(9.6)\times 10^{-4}$ & $1.5(1.1)\times 10^{-3}$ & $5.1\times 10^{-2}$ & $5.2\times 10^{-2}$ & $5.3\times 10^{-2}$ & $5.7\times 10^{-2}$\\
 $|\eta_{13}^L|<$  & $3.1(2.7)\times 10^{-2}$ & $3.3\times 10^{-2}$ & $1.1\times 10^{-2}$ & $1.0\times 10^{-2}$ & $4.8\times 10^{-2}$ & $4.8\times 10^{-2}$\\
 $|\eta_{23}^L|<$  & $2.8(2.2)\times 10^{-2}$ & $3.0\times 10^{-2}$ & $5.5\times 10^{-2}$ & $5.4\times 10^{-2}$ & $1.2\times 10^{-2}$ & $1.2\times 10^{-2}$\\  \hline \hline
\end{tabular}
\caption{\label{tab-results2}  Limits on unitarity 
  violation parameters from lepton flavor violation searches. 
  The numbers given in parenthesis correspond to the improvement 
  obtained with the recent MEG limit on $\mu  \to e \gamma$. Other 
  entries in the table are unchanged. These limits express the
  correlation between lepton non-unitarity and LFV that holds in 
  low-scale seesaw schemes under a ``minimal flavor violation hypothesis'' defined in the text.}
\end{table}

\vskip 1cm
Before closing let us comment on the robustness or our results with
respect to the assumptions made in Sec.~\ref{sec:numerical-analysis}.
Insofar as the regions obtained in
Figs.~(\ref{fig-inverse-br-NH-me})-(\ref{fig-linear-br-IH-tm}) are
concerned, we can state that they remain ``allowed'' once one departs
from our simplifying assumptions.
As expected however, we have verified that the regions obtained away
from the simplifying assumptions may allow for somewhat larger values
of the lepton-flavor-violating parameters $\eta$ affecting neutrino
propagation.
However, on account of weak universality constraints, we prefer to
stick to the more conservative values we have presented in
Table~\ref{tab-results2}.

\section{Conclusions}
\label{c}

The physics responsible for neutrino masses could lie at the TeV
scale. In this case it is very unlikely that neutrino masses are not
accompanied by non-standard neutrino interactions that could reveal
novel features in neutrino propagation. Similarly, lepton flavor
violation should also take place in processes involving the charged
leptons.

Within low-scale seesaw mechanisms, such as the inverse and linear
type-I seesaw, we have found that non-unitarity in the lepton mixing
matrix up to the percent level in some cases, is consistent with the
constraints that follow from lepton flavor violation searches in the
laboratory. This conclusion holds even within the simple ``minimal
flavor violation'' assumptions we have made.
As a result the upcoming long baseline neutrino
experiments~\cite{bandyopadhyay:2007kx} do provide an important window
of opportunity to perform complementary tests of lepton flavor
violation in neutrino propagation and to probe the mass scale
characterizing the seesaw mechanism.

\vfill

\section{Acknowledgments}

We thank M. Hirsch for useful discussions.  This work was supported by
the Spanish MICINN under grants FPA2008-00319/FPA and MULTIDARK
CSD2009-00064, by Prometeo/2009/091 (Generalitat valenciana), by the
EU Network grant UNILHC PITN-GA-2009-237920. S. M. is supported by a
Juan de la Cierva contract, M. T.  acknowledges financial
support from CSIC under the JAE-Doc programme.

\bibliographystyle{h-physrev}

\end{document}